\documentclass[11pt,floatfix,notitlepage,byrevtex]{article}
\usepackage{graphicx,epstopdf}
\usepackage[totalwidth=500pt, totalheight=650pt]{geometry}
\usepackage{pstricks,pst-plot,pst-grad,pst-3dplot,pstricks-add,pst-node}
\usepackage{mathrsfs}   
\usepackage{amsmath}    
\usepackage{amssymb}    
\usepackage{CJKutf8,CJKnumb}
\usepackage{hyperref}
\usepackage{CJK}
\usepackage{amsfonts}
\usepackage{caption2,graphics,placeins,verbatim}
\usepackage{diagbox}
\usepackage[numbers,sort&compress]{natbib}
\usepackage{authblk}

\allowdisplaybreaks

\title{Effective field theories on subspaces of the Bruhat-Tits tree}
\author{Feng Qu\thanks{qufeng@syu.edu.cn}}
\affil{\small{The Normal College, Shenyang University, Shenyang, P.~R.~China}}
\date{}

\begin{document}
\maketitle

\begin{abstract}
On two subspaces of the Bruhat-Tits tree, effective actions are calculated. The limits of these effective field theories are found to be the same conformal field theory over p-adic numbers when subspaces are taken to the boundary of the tree. Their relations to the p-adic version of AdS/CFT are also discussed.
\end{abstract}

\section{Introduction}

There are at least two motivations of studying physics over p-adic numbers $\mathbb{Q}_p$. One comes from the possibility that spacetime is non-Archimedean under small scales~\cite{IVVolovich_1987}, and the other relates to the "number field invariance principle"~\cite{Volovich:1987wu}. This subject begins with p-adic strings~\cite{FREUND1987186,FREUND1987191,IVVolovich_1987,BREKKE1988365,Zabrodin:1988ep}, and lots of works follow such as those on gravity~\cite{Dimitrijevic:2015eaa,Gubser:2016htz,Huang:2019nog,Chen:2021rsy,Dragovich:2022vvh}, on the anti-de Sitter/conformal field theory correspondence (AdS/CFT)~\cite{Gubser:2016guj,Heydeman:2016ldy,Bhattacharyya:2017aly,Hung:2019zsk,Qu:2021fgz} and on spinor~\cite{Gubser:2018cha,Qu:2019tyi,Garcia-Compean:2022pjw}. Please refer to~\cite{Vladimirov:1994wi} for useful knowledge on $\mathbb{Q}_p$.

\begin{figure}[tbp]
	\centering
	\includegraphics[width=0.7\textwidth]{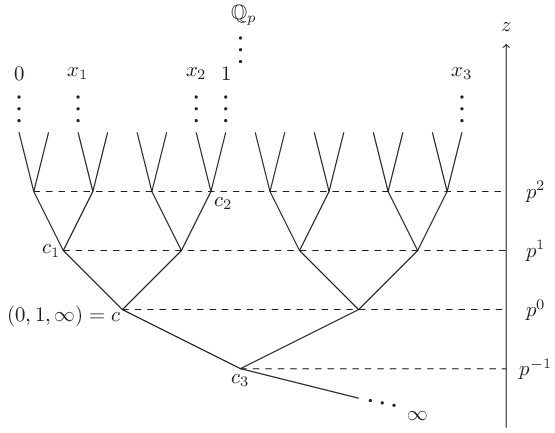}
	\caption{\label{Fig:bttree}The Bruhat-tits tree $\textrm{T}_p$ with the prime number $p=2$. It is an infinite tree with $p+1$ neighbors for each vertex, and the boundary is $\mathbb{Q}_p\cup\{\infty\}$. The p-adic absolute value $|\cdot|_p$ can be determined according to the $z$-coordinate. For example, $|x_1-x_2|_p=|z(c)|_p=|p^0|_p=p^{-0}=1$, where $c$ is the lowest vertex on the line connecting $x_1$ and $x_2$ which is denoted as $\overline{x_1x_2}$. Similarly we have $|x_3|_p=|x_3-0|_p=p^1$, $|x_2-1|_p=p^{-2}$. $c$ is the common vertex of $\overline{0\infty}~,~\overline{01}$ and $\overline{1\infty}$, hence $c=c(0,1,\infty)$.}
\end{figure}

\begin{figure}[tbp]
	\centering
	\includegraphics[width=0.7\textwidth]{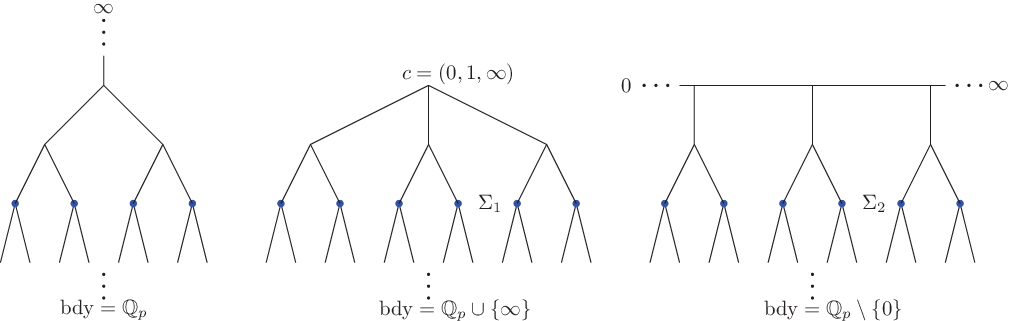}
	\caption{\label{Fig:subspace}Three subspaces (blue vertices) of $\textrm{T}_{p=2}$. The middle and right ones are denoted as $\Sigma_1$ and $\Sigma_2$.}
\end{figure}

This paper is devoted to some exact results of effective field theories over $\mathbb{Q}_p$. Referring to figure~\ref{Fig:bttree}, the Bruhat-Tits tree ($\textrm{T}_p$) can be regarded as the AdS space over $\mathbb{Q}_p$~\cite{Gubser:2016guj}. Please refer to~\cite{Guilloux_2016,Qu:2021huo} for examples of hyperboloids over $\mathbb{Q}_p$. Three subspaces of $\textrm{T}_p$ are presented in figure~\ref{Fig:subspace}, and effective action on the left one has been studied in our previous work~\cite{Qu:2021fgz}. Effective actions on the other two subspaces ($\Sigma_1$ and $\Sigma_2$) are calculated in this paper by integrating out fields on $\textrm{T}_p\setminus\Sigma_i~,~i=1,2$ where "$\textrm{T}_p\setminus\Sigma_i$" means "$\textrm{T}_p$ minus $\Sigma_i$" or "the complement of $\Sigma_i$".

The key result of AdS/CFT~\cite{Maldacena:1997re,Gubser:1998bc,Witten:1998qj} can be written as
\begin{gather}
\langle e^{\int dxO\phi_{0}}\rangle_{\textrm{CFT}}=\int_{\textrm{bulk}}\mathcal{D}\phi e^{-S[\phi]}\Big|_{\phi(\textrm{bdy})=\phi_{0}}~.
\end{gather}
The operator $O$ is coupled to a source $\phi_{0}$, and $\phi$ is integrated out on the bulk of AdS with its boundary value $\phi(\textrm{bdy})=\phi_{0}$. Please refer to~\cite{Grado-White:2020wlb,Faulkner:2010jy,McGough:2016lol} for deformed AdS/CFT. If rewriting the above equation as
\begin{gather}\label{eqn:padscft}
\langle e^{\int dxO\phi_{0}}\rangle_{\textrm{CFT}}=\lim_{\Sigma\to\partial\textrm{AdS}}\int_{\textrm{AdS}\setminus\Sigma}\mathcal{D}\phi e^{-S[\phi]}\Big|_{\phi(\Sigma)=\phi_{0}}=\lim_{\Sigma\to\partial\textrm{AdS}}e^{-S_{\Sigma}[\phi_0]}~,
\end{gather}
where "$\Sigma\to\partial\textrm{AdS}$" means "the subspace $\Sigma$ tends to the boundary of AdS" and "$S_{\Sigma}$" represents the effective action on $\Sigma$, we can relate our work of this paper to the p-adic version of AdS/CFT. 

The discussion on the only CFT over $\mathbb{Q}_p$ which has an explicit action can be found in~\cite{Melzer:1989IJMPA}, and the action is written as
\begin{gather}\label{eqn:pcft}
S_{p\textrm{CFT}}\propto\int_{\mathbb{Q}_p}dx\int_{\mathbb{Q}_p}dy\frac{\phi(x)\phi(y)}{|x-y|_p^2}\sim\int_{\mathbb{Q}_p}dx\int_{\mathbb{Q}_p}dy\frac{\phi(x)(\phi(x)-\phi(y))}{|x-y|_p^2}~,
\end{gather}
where $dx$ is the invariant measure under $x\to x+a~,~a\in\mathbb{Q}_p$. The expression behind "$\sim$" can be regarded as the regularized version of the middle one~\cite{Gubser:2016htz}. It is worth mentioning that $S_{p\textrm{CFT}}$ is actually the limit of the effective field theory in our previous work~\cite{Qu:2021fgz} when the subspace is taken to the boundary of $\textrm{T}_p$, and the same situation also occurs in this paper. An example of two-point functions is given as
\begin{gather}
V_{\alpha}(x)=e^{i\alpha\phi(x)}~,~\alpha\in\mathbb{R}~,
\\
\langle V_{\alpha}(x)V_{-\alpha}(y)\rangle=|x-y|_p^{-2\alpha^2}~.
\end{gather}

The structure of this paper is as follows. Section~\ref{sec:action} is the preparation of the action. Effective actions on subspaces and the boundary of $\textrm{T}_p$ are calculated in section~\ref{sec:eft} and~\ref{sec:bdyoftp}. The relations to p-adic AdS/CFT are discussed in section~\ref{sec:twop}. The last one is the summary and discussion. We set p-adic numbers and theirs p-adic absolute values dimensionless in this paper.

\section{The Action on $\textrm{T}_p$}\label{sec:action}

Consider a free massless scalar field on vertices of $\textrm{T}_{p}$. The action can be written as
\begin{gather}\label{eqn:actionofscalar}
S=\frac{1}{2}\sum_{\langle ab\rangle}\frac{(\phi_a-\phi_b)^2}{L_0^2}=\frac{1}{4L_0^2}\sum_a\sum_{b\in\partial a}(\phi_a-\phi_b)^2~.
\end{gather}
$\langle ab\rangle$ means the sum is over all edges of $\textrm{T}_p$ whose endpoints are denoted as $a$ and $b$, and $b\in\partial a$ means the sum is over all neighboring vertices of the given vertex $a$. $L_0$ is the length of each edge, which is a constant with a dimension of length.

\begin{figure}[tbp]
	\centering
	\includegraphics[width=0.7\textwidth]{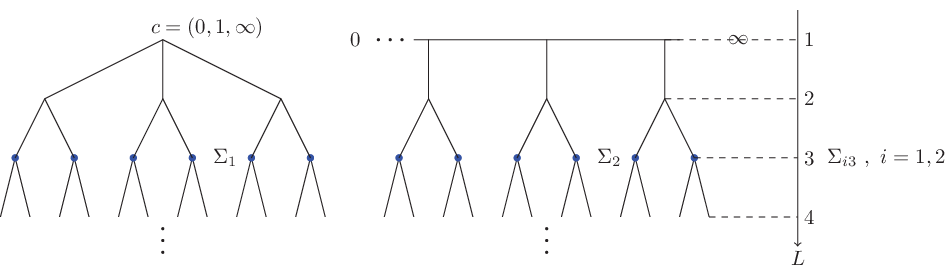}
	\caption{\label{Fig:l}$L-$coordinate on $\textrm{T}_{p=2}$. Blue vertices identify subspaces $\Sigma_1$ and $\Sigma_2$ at $L=3$. $\Sigma_{i}~,~i=1,2$ at $L=N$ can be denoted as $\Sigma_{iN}$.}
\end{figure}

Referring to figure~\ref{Fig:l}, let's introduce an $L-$coordinate on $\textrm{T}_p$. On the left in figure~\ref{Fig:l}, the $L-$coordinate of a vertex $a$, namely $L(a)$, represents the distance (number of edges) between $a$ and vertex $c=(0,1,\infty)$. To be exact, the distance between $a$ and $c$ are given by $L(a)-L(c)=L(a)-1$. On the right in figure~\ref{Fig:l}, the distance between a vertex $a$ and line $\overline{0\infty}$ are given by $L(a)-1$. The action can be rewritten as "field times EOM (equation of motion)" which is
\begin{gather}
S=\frac{1}{2L_0^2}\sum_{L(a)\leq M}\phi_a\Box\phi_a+\frac{1}{4L_0^2}R_M(\phi,\phi)+\frac{1}{4L_0^2}F_M(\phi,\phi)~,
\\
\Box f_a:=\sum_{b\in\partial a}(f_a-f_b)~,
\\
R_M(f,g):=\sum_{L(a)>M}\sum_{b\in\partial a}(f_a-f_b)(g_a-g_b)~,
\\
F_M(f,g):=\sum_{L(a)=M}\sum_{\substack{b\in\partial a\\L(b)=M+1}}(f_a+f_b)(g_b-g_a)~.
\end{gather}
$M$ is a positive integer. $f$ and $g$ represent any two fields on $\textrm{T}_p$, and "$:=$" means "be defined as". For simplicity, we only consider fields or the field space satisfying $R_M,F_M\to0$ when $M\to+\infty$, hence the action can also be written as 
\begin{gather}
S=\frac{1}{2L_0^2}\sum_{a}\phi_a\Box\phi_a~.
\end{gather}
The sum is over all vertices of $\textrm{T}_p$. Furthermore, in this field space, it can be deduced that 
\begin{gather}\label{eqn:bdyeom}
\sum_{a}f_a\Box g_a=\sum_{a}g_a\Box f_a~.
\end{gather}

Denote $\Sigma_1$ and $\Sigma_2$ at $L=N$ as $\Sigma_{iN}$ where $i=1,2$. There is an example for $\Sigma_{i3}$ in figure~\ref{Fig:l}. Decompose $\phi$ into $\Phi$ which is on-shell on $\textrm{T}_p\setminus\Sigma_{iN}$, namely the complement of $\Sigma_{iN}$, and $\phi'$ which vanishes on $\Sigma_{iN}$
\begin{gather}
\phi_a=\Phi_a+\phi'_a~,~\Box\Phi_a|_{a\notin\Sigma_{iN}}=0~,~\phi'_a|_{a\in\Sigma_{iN}}=0~.
\end{gather}
With the help of (\ref{eqn:bdyeom}), for $N\geq2$ the action can be written as
\begin{gather}
S=\frac{1}{2L_0^2}\sum_{a\in\Sigma_{iN}}\Phi_a(\Phi_a-\Phi_{a^-})+\frac{1}{2L_0^2}\sum_{a\in\Sigma_{iN}}\Phi_a(p\Phi_a-\sum_{\substack{b\in\partial a\\L(b)>L(a)}}\Phi_b)+S'~,
\\
S'=\frac{1}{2L_0^2}\sum_{a\in\textrm{T}_p\setminus\Sigma_{iN}}\phi'_a\Box\phi'_a~,
\end{gather}
where $a^-$ is the neighboring vertex of $a$ satisfying $L(a^-)+1=L(a)=N$. We can choose a particular on-shell configuration of $\Phi_b$ as
\begin{gather}
\Phi_b=\Phi_{b^-}~\textrm{when}~L(b)>L(a)~,
\end{gather}
where $b^-$ denotes the neighboring vertex of $b$ satisfying $L(b^-)+1=L(b)$. It is worth mentioning that this configuration satisfies the Neumann boundary condition $\partial\Phi(x)=0$ on the boundary of $\textrm{T}_p$~\cite{Zabrodin:1988ep}, and the derivative on the boundary is defined as
\begin{gather}
\partial f(x)\propto\lim_{a\to x}\frac{f(x)-f(a)}{p^{-L(a)}}~,
\end{gather}
where "$a\to x$" means "the vertex $a$ tends to the boundary point $x$". As for the second term in the action, we have
\begin{gather}
\begin{aligned}
&\frac{1}{2L_0^2}\sum_{a\in\sum_{iN}}\Phi_a(p\Phi_a-\sum_{\substack{b\in\partial a\\L(b)>L(a)}}\Phi_b)=\frac{1}{2L_0^2}\sum_{a\in\sum_{iN}}\Phi_a(p\Phi_a-\sum_{\substack{b\in\partial a\\L(b)>L(a)}}\Phi_a)=0~.
\end{aligned}
\end{gather}
And the action can be rewritten as
\begin{gather}\label{eqn:incompleteaction}
S=\frac{1}{2L_0^2}\sum_{a\in\Sigma_{iN}}\Phi_a(\Phi_a-\Phi_{a^-})+S'=S_{iN}+S'~.
\end{gather}
If the on-shell $\Phi_{a^-}$ can be reconstructed from $\Phi$'s on $\Sigma_{iN}$, $S_{iN}$ turns to be an action on $\Sigma_{iN}$. It is found in the end of the next section that $S_{iN}$ is the effective action.

\section{Field Reconstructions and Effective Actions on Subspaces}\label{sec:eft}

\begin{figure}[tbp]
	\centering
	\includegraphics[width=0.65\textwidth]{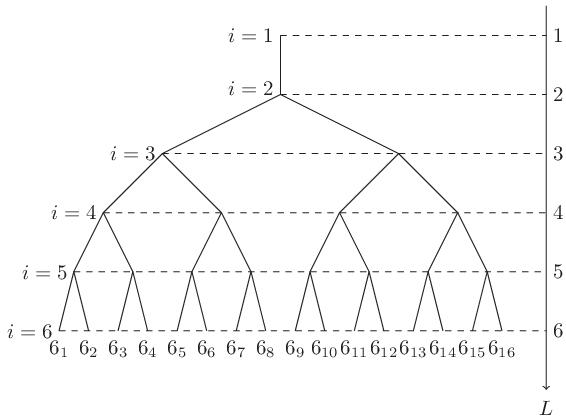}
	\caption{\label{Fig:twoeqn}One subgraph of $\textrm{T}_{p=2}$ with $5$ edges high. The index $i=1,2,\cdots,6$ denotes a vertex at $L=i$, and there are more than one vertex except $i=1,2$. For example, $\phi_{6}$ (the field at $L=6$) can represent any one of $\phi_{6_1},\phi_{6_2},\phi_{6_3},\cdots,\phi_{6_{15}},\phi_{6_{16}}$. Vertices $1$ and $6_j~,~j=1,2,\cdots,15,16$ are boundary points of the subgraph.}
\end{figure}

There are two equations very useful for the field reconstruction. Consider a special kind of subgraph of $\textrm{T}_p$ as shown in figure~\ref{Fig:twoeqn}. Image a similar subgraph which is $N-1\geq2$ edges high, and there are vertices at $L=1,2,\cdots,N-1,N$. Boundary points are located at $L=1,N$. Setting the free massless scalar field inside the boundary ($L=2,\cdots,N-1$) on-shell, the first useful equation is the reconstruction of $\phi_2$ from $\phi_1$ and $\phi_N$'s, and it can be written as
\begin{gather}\label{eqn:1eqn}
\phi_2=\frac{p^{N-2}-1}{p^{N-1}-1}\phi_{1}+\frac{p-1}{p^{N-1}-1}\sum_{N_i\in2}\phi_{N_i}~,~N\geq3~.
\end{gather}
$\phi_{N_i}$ represents the field on the lower boundary ($L=N$). $N_i\in2$ means the sum is over all vertices at $L=N$ those connect to vertex $2$ from below. For example in figure~\ref{Fig:twoeqn}, we have
\begin{gather}
6_1,6_2\in5~,~6_1,6_2,6_3,6_4\in4~,~6_1,\cdots,6_8\in3~,~6_1,\cdots,6_{16}\in2~.
\end{gather}
It can be found that $N_i\in2$ also means the sum is over all fields on the lower boundary. The equation (\ref{eqn:1eqn}) can be proved by mathematical induction.

The second useful equation is the reconstruction of $\phi_{N-1}$ from $\phi_1$ and $\phi_N$'s. Replacing $(\phi_2,\phi_1,p^N,N_i\in2)$ in (\ref{eqn:1eqn}) with $(\phi_2,\phi_1,p^N,N_i\in2),(\phi_3,\phi_2,p^{N-1},N_i\in3),(\phi_4,\phi_3,p^{N-2},N_i\in4),\cdots,(\phi_{N-1},\phi_{N-2},p^3,N_i\in{N-1})$ gives a set of equations. After eliminating $\phi_2,\phi_3,\cdots,\phi_{N-2}$ and introducing a resummation, the reconstruction of $\phi_{N-1}$ can be written as
\begin{gather}
\begin{aligned}\label{eqn:2eqn}
\phi_{N-1}=&\frac{1}{S_{N-1}}\phi_{1}
\\
&+\Big(\frac{1}{S_1S_{2}}+\frac{1}{S_2S_{3}}+\cdots+\frac{1}{S_{N-2}S_{N-1}}\Big)\sum_{d=2\times1-1}\phi_{N_i}
\\
&+\Big(\frac{1}{S_2S_{3}}+\cdots+\frac{1}{S_{N-2}S_{N-1}}\Big)\sum_{d=2\times2-1}\phi_{N_i}
\\
&+\cdots
\\
&+\frac{1}{S_{N-2}S_{N-1}}\sum_{d=2\times(N-2)-1}\phi_{N_i}~,~N\geq3~,
\end{aligned}
\\
S_k:=1+p+p^2+\cdots+p^{k-1}=\frac{p^k-1}{p-1}~.
\end{gather}
$d$ is the distance (number of edges) between the boundary vertex at $L=N$ and the location of $\phi_{N-1}$.

\begin{figure}[tbp]
	\centering
	\includegraphics[width=0.8\textwidth]{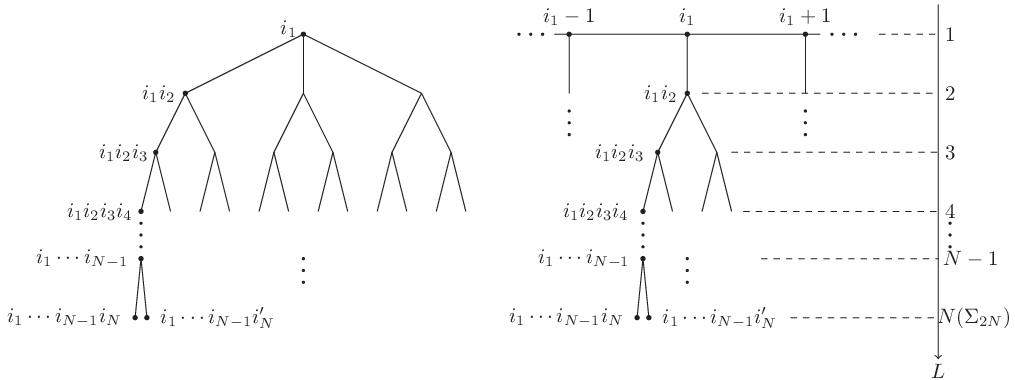}
	\caption{\label{Fig:recon}Notations of vertices for the field reconstruction on $\textrm{T}_{p=2}$. Vertices at $L=1$ are denoted as $i_1$, and there is only one vertex $i_1$ on the left. On the right there are infinite vertices at $L=1$, which are denoted as $-\infty,\cdots,i_1-2,i_1-1,i_1,i_1+1,i_1+2,\cdots,+\infty$, and vertices at $L=k$ are denoted as $i_1i_2i_3\cdots i_k$ where $i_1=-\infty,\cdots,+\infty;~i_2=1,2,\cdots,p-1;~i_3=1,2,\cdots,p;~\cdots;~i_k=1,2,\cdots,p$. Subspace $\Sigma_{2N}$ is located at $L=N$. $i_1\cdots i_{N-1}i_{N}$ and $i_1\cdots i_{N-1}i_{N}'$ are two different vertices belonging to the same vertex $i_1\cdots i_{N-1}$ when $i_N\neq i_N'$.}
\end{figure}

As for the reconstruction of $\Phi_{a^-}$ in (\ref{eqn:incompleteaction}), take the case of $\Sigma_2$($\Sigma_{2N}$) in figure~\ref{Fig:l} as an example. Please refer to the right in figure~\ref{Fig:recon} for notations.
According to (\ref{eqn:2eqn}), $\Phi_{i_1i_2\cdots i_{N-1}}~,~N\geq3$ can be reconstructed from $\Phi_{i_1}$ and $\Phi_{i_1'i_2'\cdots i_{N-1}'i_N'}$, and the remaining question is reconstructing $\Phi_{i_1}$ from $\Phi_{i_1'\cdots i_{N}'}$. According to (\ref{eqn:1eqn}), $\Phi_{i_1i_2}$ can be reconstructed from $\Phi_{i_1}$ and $\Phi_{i_1'\cdots i_{N}'}$. Summing over $i_2$ and using EOM on $i_1$ lead to
\begin{gather}
\left\{
\begin{aligned}
\sum_{i_2}\Phi_{i_1i_2}=&(p-1)\frac{p^{N-2}-1}{p^{N-1}-1}\Phi_{i_1}+\frac{p-1}{p^{N-1}-1}\sum_{i_2}\sum_{i_1'\cdots i_{N}'\in i_1i_2}\Phi_{i_1'\cdots i_{N}'}
\\
(p+1)\Phi_{i_1}=&\Phi_{i_1-1}+\Phi_{i_1+1}+\sum_{i_2}\Phi_{i_1i_2}
\end{aligned}
\right.~.
\end{gather}
Considering that $\sum_{i_2}\sum_{i_1'\cdots i_{N}'\in i_1i_2}=\sum_{i_1'\cdots i_{N}'\in i_1}$, eliminating $\sum_{i_2}\Phi_{i_1i_2}$ leads to
\begin{gather}\label{eqn:cc0}
\Phi_{i_1-1}+c\Phi_{i_1}+\Phi_{i_1+1}=-\frac{p-1}{p^{N-1}-1}\sum_{i_1'\cdots i_{N}'\in i_1}\Phi_{i_1'\cdots i_{N}'}~,~c=\frac{p^N+p^{N-2}-2}{1-p^{N-1}}~,
\end{gather}
which is also correct for $N=2$. The reconstruction of $\Phi_{i_1}$ from $\Phi_{i_1'\cdots i_{N}'}$ requires to find the inverse of the matrix
\begin{gather}
C=
\begin{pmatrix}
\ddots&&&&&&&&
\\
&c&1&&&&&&
\\
&1&c&1&&&&&
\\
&&1&c&1&&&&
\\
&&&1&c&1&&&
\\
&&&&1&c&1&&
\\
&&&&&1&c&1&
\\
&&&&&&1&c&
\\
&&&&&&&&\ddots
\end{pmatrix}~.
\end{gather} 
And it can be solved by the ansatz
\begin{gather}
C^{-1}=c'
\begin{pmatrix}
\ddots&&&&&&&&
\\
&1&e^{-1}&e^{-2}&e^{-3}&e^{-4}&e^{-5}&e^{-6}&
\\
&e^{-1}&1&e^{-1}&e^{-2}&e^{-3}&e^{-4}&e^{-5}&
\\
&e^{-2}&e^{-1}&1&e^{-1}&e^{-2}&e^{-3}&e^{-4}&
\\
&e^{-3}&e^{-2}&e^{-1}&1&e^{-1}&e^{-2}&e^{-3}&
\\
&e^{-4}&e^{-3}&e^{-2}&e^{-1}&1&e^{-1}&e^{-2}&
\\
&e^{-5}&e^{-4}&e^{-3}&e^{-2}&e^{-1}&1&e^{-1}&
\\
&e^{-6}&e^{-5}&e^{-4}&e^{-3}&e^{-2}&e^{-1}&1&
\\
&&&&&&&&\ddots
\end{pmatrix}~.
\end{gather}
$CC^{-1}=1$ demands that 
\begin{gather}
c'=\frac{1}{-\sqrt{c^2-4}}~,~e=\frac{-c+\sqrt{c^2-4}}{2}~.
\end{gather}
It can be checked that $c$ in (\ref{eqn:cc0}) satisfies $c<-2$, and that leads to $e>1$. With the help of $C^{-1}$, $\Phi_{i_1}$ can be reconstructed from $\Phi_{i_1'\cdots i_{N}'}$. After a resummation, it can be written as
\begin{gather}\label{eqn:recon2}
\Phi_{i_1}=\frac{1}{S_{N-1}\sqrt{c^2-4}}\Big(\sum_{i_1'\cdots i_{N}'\in i_1}\Phi_{i_1'\cdots i_{N}'}+\sum_{k=1}^{+\infty}e^{-k}\sum_{d=2\times(N-1)-1+k}\Phi_{i_1'\cdots i_{N}'}\Big)~.
\end{gather}
$d$ is the distance between $i_{1}'\cdots i_{N}'$ and $i_1\cdots i_{N-1}$. At last, substituting (\ref{eqn:recon2}) into the reconstruction of $\Phi_{i_1\cdots i_{N-1}}$ from $\Phi_{i_{1}}$ and $\Phi_{i_{1}'\cdots i_{N}'}$, the reconstruction of $\Phi_{i_1\cdots i_{N-1}}$ can be written as
\begin{gather}\label{eqn:recon3}
\begin{aligned}
\Phi_{i_1\cdots i_{N-1}}=&\Big(\frac{1}{S_1S_{2}}+\frac{1}{S_2S_{3}}+\cdots+\frac{1}{S_{N-2}S_{N-1}}+\frac{1}{S_{N-1}^2\sqrt{c^2-4}}\Big)\sum_{d=2\times1-1}\Phi_{i_1'\cdots i_{N}'}
\\
&+\Big(\frac{1}{S_2S_{3}}+\cdots+\frac{1}{S_{N-2}S_{N-1}}+\frac{1}{S_{N-1}^2\sqrt{c^2-4}}\Big)\sum_{d=2\times2-1}\Phi_{i_1'\cdots i_{N}'}
\\
&+\cdots
\\
&+\Big(\frac{1}{S_{N-2}S_{N-1}}+\frac{1}{S_{N-1}^2\sqrt{c^2-4}}\Big)\sum_{d=2\times(N-2)-1}\Phi_{i_1'\cdots i_{N}'}
\\
&+\frac{1}{S_{N-1}^2\sqrt{c^2-4}}\sum_{d=2\times(N-1)-1}\Phi_{i_1'\cdots i_{N}'}
\\
&+\frac{1}{S_{N-1}^2\sqrt{c^2-4}}\sum_{k=1}^{+\infty}e^{-k}\sum_{d=2\times(N-1)-1+k}\Phi_{i_1'\cdots i_{N}'}~,
\end{aligned}
\end{gather}
which is also correct for $N=2$, hence $N\geq2$. The field reconstruction in the case of subspace $\Sigma_{1N}$ is easier, and the result can be written as
\begin{gather}\label{eqn:recon4}
\begin{aligned}
\Phi_{i_1\cdots i_{N-1}}=&\Big(\frac{1}{S_1S_{2}}+\frac{1}{S_2S_{3}}+\cdots+\frac{1}{S_{N-2}S_{N-1}}+\frac{1}{S_{N-1}S_2p^{N-2}}\Big)\sum_{d=2\times1-1}\Phi_{i_1'\cdots i_{N}'}
\\
&+\Big(\frac{1}{S_2S_{3}}+\cdots+\frac{1}{S_{N-2}S_{N-1}}+\frac{1}{S_{N-1}S_2p^{N-2}}\Big)\sum_{d=2\times2-1}\Phi_{i_1'\cdots i_{N}'}
\\
&+\cdots
\\
&+\Big(\frac{1}{S_{N-2}S_{N-1}}+\frac{1}{S_{N-1}S_2p^{N-2}}\Big)\sum_{d=2\times(N-2)-1}\Phi_{i_1'\cdots i_{N}'}
\\
&+\frac{1}{S_{N-1}S_2p^{N-2}}\sum_{d=2\times(N-1)-1}\Phi_{i_1'\cdots i_{N}'}~,~N\geq2~.
\end{aligned}
\end{gather}
The vertex notations are the same as those in the case of subspace $\Sigma_{2N}$, and they are shown on the left in figure~\ref{Fig:recon}. $d$ still denotes the distance between $i_1'\cdots i_{N}'$ and $i_1\cdots i_{N-1}$.

According to the reconstruction of $\Phi_{i_1\cdots i_{N-1}}$ in both cases (\ref{eqn:recon4}) and (\ref{eqn:recon3}), $\Phi_{a^-}$ in (\ref{eqn:incompleteaction}) can be reconstructed from $\Phi$'s at $\Sigma_{1N}$ and $\Sigma_{2N}$. Together with $S_{iN}$ in (\ref{eqn:incompleteaction}), we have
\begin{gather}
\left\{
\begin{aligned}\label{eqn:effaction1}
S_{1N}=&\frac{1}{2L_0^2}\sum_{a\in\Sigma_{1N}}\Phi_a\sum_{\substack{b\in\Sigma_{1N}\\b\neq a}}A_{\frac{d(a,b)}{2}}(\Phi_a-\Phi_b)
\\
A_n:=&\frac{1}{S_nS_{n+1}}+\frac{1}{S_{n+1}S_{n+2}}+\cdots+\frac{1}{S_{N-2}S_{N-1}}+\frac{1}{S_{N-1}S_2p^{N-2}}
\end{aligned}
\right.~,
\end{gather}
\begin{gather}
\left\{
\begin{aligned}\label{eqn:effaction2}
S_{2N}=&\frac{1}{2L_0^2}\sum_{a\in\Sigma_{2N}}\Phi_a\Big(\sum_{\substack{b\in\Sigma_{2N}\\b\neq a\\d(a,b)\leq2(N-1)}}B_{\frac{d(a,b)}{2}}(\Phi_a-\Phi_b)
\\
&+\sum_{\substack{b\in\Sigma_{2N}\\d(a,b)>2(N-1)}}C_{d(a,b)-2(N-1)}(\Phi_a-\Phi_b)\Big)
\\
B_n:=&\frac{1}{S_nS_{n+1}}+\frac{1}{S_{n+1}S_{n+2}}+\cdots+\frac{1}{S_{N-2}S_{N-1}}+\frac{1}{S_{N-1}^2\sqrt{c^2-4}}
\\
C_n:=&\frac{1}{S_{N-1}^2\sqrt{c^2-4}}e^{-n}
\end{aligned}
\right.~,
\end{gather}
where $d(a,b)$ is the distance between $a$ and $b$. The following equations have been used:
\begin{gather}
\sum_{b\in\Sigma_{1N}}A_{\frac{1+d(a^-,b)}{2}}=1~,
\\
\sum_{\substack{b\in\Sigma_{2N}\\d(a^-,b)\leq2(N-1)-1}}B_{\frac{1+d(a^-,b)}{2}}+\sum_{\substack{b\in\Sigma_{2N}\\d(a^-,b)>2(N-1)-1}}C_{1+d(a^-,b)-2(N-1)}=1~,
\\
b\neq a\Rightarrow d(a,b)=d(a^-,b)+1~.
\end{gather}

After writing $S_{iN}$ as an action on $\Sigma_{iN}$, we can find the effective action by integrating out fields on $\textrm{T}_p\setminus\Sigma_{iN}$. Introducing a source $J$ which only lives on $\Sigma_{iN}$, the partition function of $\phi$ can be written as
\begin{gather}
Z[J]=\frac{\int_{\textrm{T}_p}\mathcal{D}\phi e^{-S+\sum_{a\in\Sigma_{iN}}\phi_aJ_a}}{\int_{\textrm{T}_p}\mathcal{D}\phi e^{-S}}~,
\end{gather}
where $S$ is the action in (\ref{eqn:incompleteaction}). In the numerator and denominator, functional integrals on $\textrm{T}_p\setminus\Sigma_{iN}$ are the same, hence cancelled. Only those on $\Sigma_{iN}$ are left, and that leads to
\begin{gather}
Z[J]=\frac{\int_{\Sigma_{iN}}\mathcal{D}\Phi e^{-S_{iN}+\sum_{a\in\Sigma_{iN}}\Phi_aJ_a}}{\int_{\Sigma_{iN}}\mathcal{D}\Phi e^{-S_{iN}}}~.
\end{gather}
So $S_{iN}$ is indeed the effective action on $\Sigma_{iN}$.

\section{Effective Actions on the Boundary}\label{sec:bdyoftp}

\begin{figure}[tbp]
	\centering
	\includegraphics[width=0.6\textwidth]{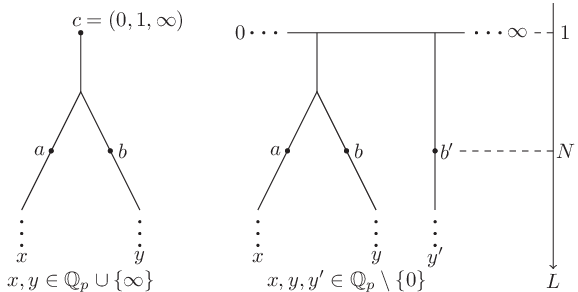}
	\caption{\label{Fig:toinfinity}The limit $N\to+\infty$. The lower infinite boundaries are $\mathbb{Q}_p\cup\{\infty\}$ on the left and $\mathbb{Q}_p\setminus\{0\}$ on the right.} 
\end{figure}

\begin{figure}[tbp]
	\centering
	\includegraphics[width=0.8\textwidth]{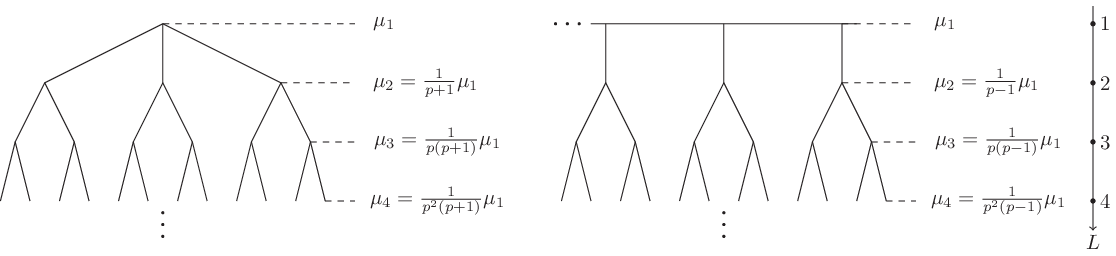}
	\caption{\label{Fig:measure}Two different measures for vertices on $\textrm{T}_{p=2}$. The measure of a vertex at $L=1$ is denoted as $\mu_1$. As for a vertex at $L=n\geq2$, the measure $\mu_n$ is set to $\mu_n=\frac{1}{p^{n-2}(p+1)}\mu_1$ on the left and $\mu_n=\frac{1}{p^{n-2}(p-1)}\mu_1$ on the right.} 
\end{figure}

Consider the case of $N\to+\infty$ when vertices at $L=N$ approach to the boundary of $\textrm{T}_p$. Referring to figure~\ref{Fig:toinfinity},
suppose that $a\to x~,~b(b')\to y(y')$ when $N\to+\infty$. When fixing boundary points $x,y(y')$, the distance between $a$ and $b(b')$ tends to infinity, namely $d(a,b),d(a,b')\to+\infty$. To found the effective action on the infinite boundary (the lower boundary in figure~\ref{Fig:toinfinity}), we need the limit behaviors of some  parameters, and it is found that
\begin{gather}
k\to+\infty:~S_k=\frac{p^k-1}{p-1}\to\frac{p^k}{p-1}~,
\\
c=\frac{p^N+p^{N-2}-2}{1-p^{N-1}}\to-p-p^{-1}~,
\\
e=\frac{-c+\sqrt{c^2-4}}{2}\to p~,
\\
A_{\frac{d(a,b)}{2}},B_{\frac{d(a,b)}{2}},C_{d(a,b)-2(N-1)}\to\frac{p(p-1)}{p+1}p^{-d(a,b)}~.\label{eqn:limitofc}
\end{gather}
It is unexpected that all three coefficients $A,B,C$ have the same limit behaviors. The effective actions in (\ref{eqn:effaction1}) for $\Sigma_{1N}$ and in (\ref{eqn:effaction2}) for $\Sigma_{2N}$ can be written in a unified form when $N\to+\infty$, which is
\begin{gather}
S_{iN}\to\frac{p(p-1)}{2(p+1)L_0^2}\sum_{a\in\Sigma_{iN}}\Phi_a\sum_{\substack{b\in\Sigma_{iN}\\b\neq a}}p^{-d(a,b)}(\Phi_a-\Phi_b)~.
\end{gather} 
$\sum$ should be replaced by $\int$ under this limit, and there have to be measure parts which tend to zero. We introduce measures for both cases as shown in figure~\ref{Fig:measure}. Adding one $\mu_N$ behind each $\sum$, effective actions can be written as
\begin{gather}
S_{1N}\to\frac{p^2-1}{2p^3\mu_1^2L_0^2}\sum_{a\in\Sigma_{1N}}\mu_N\Phi_a\sum_{\substack{b\in\Sigma_{1N}\\b\neq a}}\mu_Np^{2N-d(a,b)}(\Phi_a-\Phi_b)~,\label{eqn:toinfinity1}
\\
S_{2N}\to\frac{(p-1)^3}{2p^3(p+1)\mu_1^2L_0^2}\sum_{a\in\Sigma_{2N}}\mu_N\Phi_a\sum_{\substack{b\in\Sigma_{2N}\\b\neq a}}\mu_Np^{2N-d(a,b)}(\Phi_a-\Phi_b)~,\label{eqn:toinfinity2}
\end{gather}
where $\sum\mu_N\to\int dx$ has a dimension of length. The remaining question is to find the limit of $p^{2N-d}$. 

\begin{figure}[tbp]
	\centering
	\includegraphics[width=0.7\textwidth]{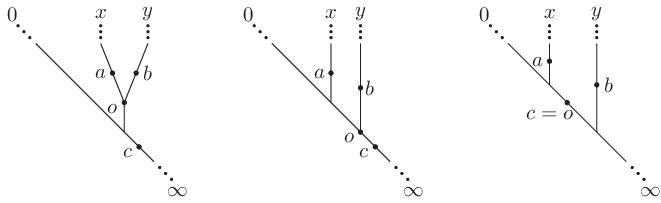}
	\caption{\label{Fig:dis1}Different position relations between $x,y$ and $c=(0,1,\infty)$. Left: $|x|_p=|y|_p\leq1$. Middle: $|x|_p<|y|_p\leq1$. Right: $|x|_p\leq1<|y|_p$. The left and middle ones correspond to the case of $|x|_p\leq|y|_p\leq1$. The case of $1<|x|_p\leq|y|_p$ is not shown in this figure. The vertex $o$ is the nearest vertex to $c$ on line $\overline{xy}$.} 
\end{figure}

Supposing that $|x|_p\leq|y|_p$, it can be classified into three cases for $\Sigma_{1N}$: $|x|_p\leq|y|_p\leq1$, $|x|_p\leq1<|y|_p$ and $1<|x|_p\leq|y|_p$. Refer to the left and middle ones in figure~\ref{Fig:dis1} for $|x|_p\leq|y|_p\leq1$. Comparing with figure~\ref{Fig:bttree} and the left one in figure~\ref{Fig:toinfinity}, it can be found that
\begin{gather}
\left.
\begin{aligned}
&L(a)=L(b)=N~,~L(c)=1
\\
&\frac{d(a,b)}{2}=d(a,o)=d(b,o)=L(b)-L(o)=N-L(o)
\\
&|x-y|_p=|z(o)|_p=|p^{L(o)-1}|_p=p^{1-L(o)}=p^{1+\frac{d(a,b)}{2}-N}
\end{aligned}
\right\}
\Rightarrow p^{2N-d(a,b)}=\frac{p^2}{|x-y|_p^2}~.\label{eqn:dis11}
\end{gather}
Referring to the right one in figure~\ref{Fig:dis1}, for $|x|_p\leq1<|y|_p$ it can be found that
\begin{gather}
\frac{d(a,b)}{2}=d(a,c)=N-1\Rightarrow p^{2N-d(a,b)}=p^2~.\label{eqn:dis12}
\end{gather}
As the case of $1<|x|_p\leq|y|_p$, applying a transformation on the boundary of $\textrm{T}_p$: $x\to x'=\frac{1}{x}$ which is an isometric transformation~\cite{Gubser:2016guj} and a rotation around $c=(0,1,\infty)$ of $\textrm{T}_p$ keeping the distance $d(a,b)$ and the $L$-coordinate $N$ in (\ref{eqn:toinfinity1}) invariant leads to the case of $|y'|_p\leq|x'|_p<1$. And that has been solved in (\ref{eqn:dis11}) already. So we can write
\begin{gather}
p^{2N-d(a,b)}=\frac{p^2}{\Big|\frac{1}{y}-\frac{1}{x}\Big|_p^2}~.\label{eqn:dis13}
\end{gather}
Consider the case of $p\equiv3\pmod4$ when $-1$ has no square root in $\mathbb{Q}_p$. The identity $|x^2+y^2|_p=|x^2,y^2|_s$~\cite{Vladimirov:1994wi} can be used to write (\ref{eqn:dis11}), (\ref{eqn:dis12}) and (\ref{eqn:dis13}) into the same form. Here $|x^2,y^2|_s$ gives the maximal one between $|x^2|_p$ and $|y^2|_p$. It can be verified that
\begin{gather}\label{eqn:dis1}
p^{2N-d(a,b)}=p^2\frac{|1+x^2|_p|1+y^2|_p}{|x-y|_p^2}~.
\end{gather}
We set $p\equiv3\pmod4$ here and below.

\begin{figure}[tbp]
	\centering
	\includegraphics[width=0.5\textwidth]{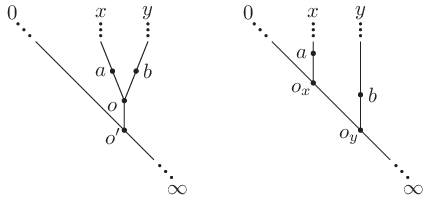}
	\caption{\label{Fig:dis2}Different position relations between $x,y$ and $\overline{0\infty}$. Left: $|x|_p=|y|_p$. Right: $|x|_p<|y|_p$.} 
\end{figure}

In the case of $\Sigma_{2N}$, still supposing that $|x|_p\leq|y|_p$, there are two different position relations between $x,y$ and $\overline{0\infty}$ which are shown in figure~\ref{Fig:dis2}.
Referring to figure~\ref{Fig:bttree}, the left in figure~\ref{Fig:dis2} and the right in figure~\ref{Fig:toinfinity}, for $|x|_p=|y|_p$ we have
\begin{gather}
\left.
\begin{aligned}
&\frac{d(a,b)}{2}=d(a,o)=L(a)-L(o)=N-L(o)
\\
&|x|_p=|y|_p~,~\frac{|x|_p}{|x-y|_p}=\frac{|z(o')|_p}{|z(o)|_p}=p^{d(o,o')}=p^{L(o)-1}
\end{aligned}\nonumber
\right\}
\\
\Rightarrow p^{2N-d(a,b)}=p^2\frac{|x|_p^2}{|x-y|_p^2}=p^2\frac{|xy|_p}{|x-y|_p^2}~.
\end{gather}
Referring to the right in figure~\ref{Fig:dis2}, for $|x|_p<|y|_p$ we have
\begin{gather}
\left.
\begin{aligned}
&d(a,b)=d(a,o_x)+d(b,o_y)+d(o_x,o_y)=2(N-1)+d(o_x,o_y)
\\
&\frac{|y|_p}{|x|_p}=\frac{|z(o_y)|_p}{|z(o_x)|_p}=p^{d(o_x,o_y)}~,~|y|_p=|x-y|_p
\end{aligned}\nonumber
\right\}
\\
\Rightarrow p^{2N-d(a,b)}=p^2\frac{|x|_p}{|y|_p}=p^2\frac{|xy|_p}{|y|_p^2}=p^2\frac{|xy|_p}{|x-y|_p^2}~.
\end{gather}
So in the case of $\Sigma_{2N}$, we can always write
\begin{gather}\label{eqn:dis2}
p^{2N-d(a,b)}=p^2\frac{|xy|_p}{|x-y|_p^2}~.
\end{gather}

Finally, substituting (\ref{eqn:dis1}) for $\Sigma_{1N}$ and (\ref{eqn:dis2}) for $\Sigma_{2N}$ into (\ref{eqn:toinfinity1}) and (\ref{eqn:toinfinity2}), effective actions on infinite boundaries ($N\to+\infty$) can be written as
\begin{gather}
\left\{
\begin{aligned}
&S_{1N}\to\frac{p^2-1}{2p\mu_1^2L_0^2}\int_{x\in\mathbb{Q}_p\cup\{\infty\}}d^{(1)}x\Phi_x\int_{\substack{y\in\mathbb{Q}_p\cup\{\infty\}\\y\neq x}}d^{(1)}y\frac{\Phi_x-\Phi_y}{|x-y|_1^2}
\\
&|x-y|_1:=\sqrt{\frac{|x-y|_p^2}{|1+x^2|_p|1+y^2|_p}}
\end{aligned}\label{eqn:ntoinf1}
\right.~,
\\
\left\{
\begin{aligned}
&S_{2N}\to\frac{(p-1)^3}{2p(p+1)\mu_1^2L_0^2}\int_{x\in\mathbb{Q}_p\setminus\{0\}}d^{(2)}x\Phi_x\int_{\substack{y\in\mathbb{Q}_p\setminus\{0\}\\y\neq x}}d^{(2)}y\frac{\Phi_x-\Phi_y}{|x-y|_2^2}
\\
&|x-y|_2:=\sqrt{\frac{|x-y|_p^2}{|xy|_p}}
\end{aligned}\label{eqn:ntoinf2}
\right.~,
\end{gather}
where $\Phi_a\to\Phi_x~,~\Phi_b\to\Phi_y$ and $\Sigma_{iN}\mu_N\to\int d^{(i)}x$ have been used. 

Moreover, $S_{iN}$($N\to+\infty$) are actually the same CFT action $S_{p\textrm{CFT}}$ in (\ref{eqn:pcft}). Referring to the left in figure~\ref{Fig:measure}, the measure $d^{(1)}x$ in (\ref{eqn:ntoinf1}) which comes from $\lim_{n\to+\infty}\mu_n$ is invariant under the isometric transformation of $\textrm{T}_p$ keeping the vertex $c=(0,1,\infty)$ fixed. According to~\cite{Zabrodin:1988ep}, we can write
\begin{gather}\label{eqn:limpcft1}
d^{(1)}x\propto\left\{
\begin{aligned}
dx~,&~|x|_p\leq1
\\
\frac{1}{|x|_p^2}dx~,&~|x|_p>1
\end{aligned}
\right\}=\frac{1}{|1+x^2|_p}dx~,
\end{gather}
where $dx$ is the same measure as that in (\ref{eqn:pcft}). The condition $p\equiv3\pmod4$ is used to obtain the equality sign. Combining (\ref{eqn:ntoinf1}) and (\ref{eqn:limpcft1}), we have
\begin{gather}\label{eqn:same1}
\lim_{N\to+\infty}S_{1N}\propto\int dx\int dy\frac{\Phi_x(\Phi_x-\Phi_y)}{|x-y|_p^2}\propto S_{p\textrm{CFT}}~.
\end{gather}
It is also correct if not imposing $p\equiv3\pmod4$, in which case $|x-y|_1$ and $d^{(1)}x$ are expressed by piecewise functions. The same situation occurs for $S_{2N}$ too. Referring to the right in figure~\ref{Fig:measure}, the measure $d^{(2)}x$ in (\ref{eqn:ntoinf2}) is invariant under the isometric transformation of $\textrm{T}_p$ transforming the line $\overline{0\infty}$ to itself which contains $x\to ax~,~a\neq0$ and $x\to\frac{1}{x}$ on the boundary of $\textrm{T}_p$. It is the Haar measure of the multiplicative group $\mathbb{Q}_p^*=\mathbb{Q}_p\setminus\{0\}$ in~\cite{Vladimirov:1994wi}, and we have
\begin{gather}\label{eqn:same2}
d^{(2)}x\propto\frac{1}{|x|_p}dx\Rightarrow\lim_{N\to+\infty}S_{2N}\propto\int dx\int dy\frac{\Phi_x(\Phi_x-\Phi_y)}{|x-y|_p^2}\propto S_{p\textrm{CFT}}~.
\end{gather}
It is found that two different effective actions $S_{iN}~,~i=1,2$ tend to the same $S_{p\textrm{CFT}}$ under the limit $N\to+\infty$.

It is worth mentioning that $|x-y|_1$ in (\ref{eqn:ntoinf1}) and $|x-y|_2$ in (\ref{eqn:ntoinf2}) still have some meanings: they can be regarded as chordal distances of a circle and a hyperbola. Introducing embedding coordinates for $\Sigma_{1N}$ and $\Sigma_{2N}$ when $N\to+\infty$. In the case of $\Sigma_{1N}$ we write
\begin{gather}
x\to(X_1,X_2)~,~y\to(Y_1,Y_2):\nonumber
\\
\left\{
\begin{aligned}
&x=\frac{X_1}{L'+X_2}
\\
&X_1=\frac{2x}{1+x^2}L'~,~X_2=\frac{1-x^2}{1+x^2}L'
\end{aligned}
\right.~,~
\left\{
\begin{aligned}
&y=\frac{Y_1}{L'+Y_2}
\\
&Y_1=\frac{2y}{1+y^2}L'~,~Y_2=\frac{1-y^2}{1+y^2}L'
\end{aligned}
\right.~,\label{eqn:trans1}
\end{gather}
where $L'$ is a dimensionless p-adic number. First, it can be found that
\begin{gather}
X_1^2+X_2^2=L'^2~,~Y_1^2+Y_2^2=L'^2~.
\end{gather}
So $(X_1,X_2)$ and $(Y_1,Y_2)$ can be regarded as two points on the same circle whose radius depends on $L'$. Second, it can be verified that
\begin{gather}
|(X_1-Y_1)^2+(X_2-Y_2)^2|_p=|L'|_p^2|x-y|_1^2~,
\end{gather}
where $|4|_p=1$ has been used when $p\equiv3\pmod4$. It means $|x-y|_1$ can be regarded as the chordal distance between two points on this circle. Introducing embedding coordinates for $\Sigma_{2N}$ when $N\to+\infty$
\begin{gather}
x\to(X_1,X_2)~,~y\to(Y_1,Y_2):\nonumber
\\
\left\{
\begin{aligned}
&x=\frac{L'}{X_2-X_1}
\\
&X_1=\frac{x^2-1}{2x}L'~,~X_2=\frac{x^2+1}{2x}L'
\end{aligned}
\right.~,~
\left\{
\begin{aligned}
&y=\frac{L'}{Y_2-Y_1}
\\
&Y_1=\frac{y^2-1}{2y}L'~,~Y_2=\frac{y^2+1}{2y}L'
\end{aligned}
\right.~.\label{eqn:trans2}
\end{gather}
First, it can be found that
\begin{gather}
X_1^2-X_2^2=-L'^2~,~Y_1^2-Y_2^2=-L'^2~.
\end{gather}
So $(X_1,X_2)$ and $(Y_1,Y_2)$ can be regarded as two points on the same hyperbola. Second, it can be verified that
\begin{gather}
|(X_1-Y_1)^2-(X_2-Y_2)^2|_p=|L'|^2|x-y|_2^2~,
\end{gather}
which means $|x-y|_2$ can be regarded as the chordal distance between two points on this hyperbola.

\section{Relations to P-adic AdS/CFT}\label{sec:twop}

The p-adic version of (\ref{eqn:padscft}) can be written as
\begin{gather}\label{eqn:padscft1}
\langle e^{\int dxO\Phi}\rangle_{\textrm{pCFT}}=\lim_{\Sigma_{iN}\to\partial\textrm{T}_p}\int_{\textrm{T}_p\setminus\Sigma_{iN}}\mathcal{D}\phi e^{-S[\phi]}\Big|_{\phi(\Sigma_{iN})=\Phi}=\lim_{N\to+\infty}e^{-S_{iN}[\Phi]}~.
\end{gather}
$\lim_{N\to+\infty}e^{-S_{iN}}$ can be regarded as the generating functional of some p-adic CFT (pCFT). According to (\ref{eqn:ntoinf1}) and (\ref{eqn:ntoinf2}), two-point functions of some operator can be written as
\begin{gather}\label{eqn:distp1}
\left\{\begin{aligned}
S_{1N\to+\infty}:&~\langle O_xO_{y\neq x}\rangle=\frac{p^2-1}{p\mu_1^2L_0^2}\frac{1}{|x-y|_1^2}
\\
S_{2N\to+\infty}:&~\langle O_xO_{y\neq x}\rangle=\frac{(p-1)^3}{p(p+1)\mu_1^2L_0^2}\frac{1}{|x-y|_2^2}
\end{aligned}\right.~.
\end{gather}
Two remarks here. First, according to (\ref{eqn:same1}) and (\ref{eqn:same2}), $S_{1N\to+\infty}$ and $S_{2N\to+\infty}$ are actually the same action $S_{p\textrm{CFT}}$ written in different measures $d^{(1)}x$ and $d^{(2)}x$. So the above two-point functions perhaps come from writing the same p-adic CFT using different measures. But we are not quite sure about it. Second, although $S_{p\textrm{CFT}}$ (the limit of $S_{iN}$) is a p-adic CFT, it is not the "CFT" in "p-adic AdS/CFT", but is the source of the "CFT" in "p-adic AdS/CFT". 

According to (\ref{eqn:padscft1}), $e^{-S_{iN}}$ may be regarded as a generating functional of some deformed p-adic CFT, and $N$ or $\frac{1}{N}$ is the deformation parameter. Referring to (\ref{eqn:effaction1}) and (\ref{eqn:effaction2}), two-point functions of deformed CFTs can be written as
\begin{gather}\label{eqn:distp2}
\left\{\begin{aligned}
S_{1N}:&~\langle O_aO_{b\neq a}\rangle=\frac{1}{L_0^2}A_{\frac{d(a,b)}{2}}~,
\\
S_{2N}:&~\langle O_aO_{b\neq a}\rangle=\left\{
\begin{aligned}
\frac{1}{L_0^2}B_{\frac{d(a,b)}{2}}~,&~d(a,b)\leq2(N-1)
\\
\frac{1}{L_0^2}C_{d(a,b)-2(N-1)}~,&~d(a,b)>2(N-1)
\end{aligned}
\right.~.
\end{aligned}\right.
\end{gather}
Two remarks here. First, the existence of deformed p-adic CFT is still a hypothesis although $S_{iN}$ may provide an example since $\lim_{N\to+\infty}S_{iN}\propto S_{p\textrm{CFT}}$. Second, besides $\textrm{T}_p$, there is another AdS space over $\mathbb{Q}_p$ proposed in \cite{Gubser:2016guj} which is denoted as $p\textrm{AdS}$. Different from $\textrm{T}_p$ discussed in this paper, the bulk of $p\textrm{AdS}$ is a continuous space. So a better discussion on AdS/CFT over $\mathbb{Q}_p$ should be based on $p\textrm{AdS}$. In this paper, we only consider fields on $\textrm{T}_p$ which is a simpler case compared to that on $p\textrm{AdS}$. $\sum_{iN}$ and $\Phi$ in (\ref{eqn:padscft1}) are a discrete space and a field living on it. 

\section{Summary and Discussion}\label{sec:sandd}

Given a free massless scalar field on $\textrm{T}_p$, we calculate effective actions on two kinds of subspaces in figure~\ref{Fig:l}. At $L=N\geq2$ for $\Sigma_{1N}$ and $\Sigma_{2N}$, these actions can be written as
\begin{gather}
S_{1N}=\frac{1}{2L_0^2}\sum_{a\in\Sigma_{1N}}\Phi_a\sum_{\substack{b\in\Sigma_{1N}\\b\neq a}}A_{\frac{d(a,b)}{2}}(\Phi_a-\Phi_b)~,
\\
S_{2N}=\frac{1}{2L_0^2}\sum_{a\in\Sigma_{2N}}\Phi_a\Big(\sum_{\substack{b\in\Sigma_{2N}\\b\neq a\\d(a,b)\leq2(N-1)}}B_{\frac{d(a,b)}{2}}(\Phi_a-\Phi_b)+\sum_{\substack{b\in\Sigma_{2N}\\d(a,b)>2(N-1)}}C_{d(a,b)-2(N-1)}(\Phi_a-\Phi_b)\Big)~.
\end{gather}
$d(a,b)$ gives the distance (number of edges) between vertex $a$ and vertex $b$. Coefficients $A,B,C$ satisfy
\begin{gather}
A_n=\frac{1}{S_nS_{n+1}}+\frac{1}{S_{n+1}S_{n+2}}+\cdots+\frac{1}{S_{N-2}S_{N-1}}+\frac{1}{S_{N-1}S_2p^{N-2}}~,
\\
B_n=\frac{1}{S_nS_{n+1}}+\frac{1}{S_{n+1}S_{n+2}}+\cdots+\frac{1}{S_{N-2}S_{N-1}}+\frac{1}{S_{N-1}^2\sqrt{c^2-4}}~,
\\
C_n=\frac{1}{S_{N-1}^2\sqrt{c^2-4}}e^{-n}~,
\\
S_k=\frac{p^k-1}{p-1}~,~c=\frac{p^N+p^{N-2}-2}{1-p^{N-1}}~,~e=\frac{-c+\sqrt{c^2-4}}{2}~.
\end{gather}

When $N\to+\infty$, $\Sigma_{1N}$ and $\Sigma_{2N}$ approach the boundary of $\textrm{T}_p$. The corresponding effective actions can be written as
\begin{gather}
S_{1N}\to\frac{p^2-1}{2p\mu_1^2L_0^2}\int_{x\in\mathbb{Q}_p\cup\{\infty\}}d^{(1)}x\Phi_x\int_{\substack{y\in\mathbb{Q}_p\cup\{\infty\}\\y\neq x}}d^{(1)}y\frac{\Phi_x-\Phi_y}{|x-y|_1^2}~,
\\
S_{2N}\to\frac{(p-1)^3}{2p(p+1)\mu_1^2L_0^2}\int_{x\in\mathbb{Q}_p\setminus\{0\}}d^{(2)}x\Phi_x\int_{\substack{y\in\mathbb{Q}_p\setminus\{0\}\\y\neq x}}d^{(2)}y\frac{\Phi_x-\Phi_y}{|x-y|_2^2}~.
\end{gather}
$d^{(1)}x$ and $d^{(2)}x$ denote two different measures on the boundary of $\textrm{T}_p$ which come from $\lim_{n\to+\infty}\mu_n$ in figure~\ref{Fig:measure}. In the case of $p\equiv3\pmod4$, $|x-y|_1$ and $|x-y|_2$ can be written using p-adic absolute value $|\cdot|_p$ as
\begin{gather}
|x-y|_1=\sqrt{\frac{|x-y|_p^2}{|1+x^2|_p|1+y^2|_p}}~,~|x-y|_2=\sqrt{\frac{|x-y|_p^2}{|xy|_p}}~.
\end{gather}
It is also found that $\lim_{N\to+\infty}S_{iN}$ is actually the same p-adic CFT in (\ref{eqn:pcft}).

Introducing embedding coordinates (\ref{eqn:trans1}) and (\ref{eqn:trans2}), it is found that 
\begin{gather}
\left\{
\begin{aligned}
&x\to(X_1,X_2)~,~y\to(Y_1,Y_2)
\\
&X_1^2+X_2^2=L'^2~,~Y_1^2+Y_2^2=L'^2
\\
&|x-y|_1^2=\frac{1}{|L'|_p^2}|(X_1-Y_1)^2+(X_2-Y_2)^2|_p
\end{aligned}
\right.~,
\\
\left\{
\begin{aligned}
&x\to(X_1,X_2)~,~y\to(Y_1,Y_2)
\\
&X_1^2-X_2^2=-L'^2~,~Y_1^2-Y_2^2=-L'^2
\\
&|x-y|_2^2=\frac{1}{|L'|_p^2}|(X_1-Y_1)^2-(X_2-Y_2)^2|_p
\end{aligned}
\right.~.
\end{gather}
So $\Sigma_{1N}$ when $N\to+\infty~$ on the left in figure~\ref{Fig:l} can be regarded as a circle over p-adic numbers, and $|x-y|_1$ as the chordal distance between $x$ and $y$ on it. Similarly, $\Sigma_{2N}$ when $N\to+\infty~$ on the right in figure~\ref{Fig:l} can be regarded as a hyperbola, and $|x-y|_2$ as its chordal distance.

Relations to (deformed) p-adic AdS/CFT are also discussed including two-point functions in (\ref{eqn:distp1}) and (\ref{eqn:distp2}).

Several problems are still unsolved, for example, (i)what are results on other subspaces besides those in figure~\ref{Fig:subspace} or in the case of $p\equiv1\pmod4$? (ii)is the limit behavior of coefficients $A,B,C$ in (\ref{eqn:limitofc}) universal? (iii)the discussion on (deformed) CFT over $\mathbb{Q}_p$ is still insufficient in this paper.

\end{document}